\documentclass{PoS}
\usepackage{amsmath,amssymb}
\usepackage{braket}

\newcommand{\tr}{\operatorname{tr}}

\title{
\begin{picture}(0,0)(0,0)%
  \put(420,180){\makebox(0,0)[l]{\textnormal{\normalsize
  CHIBA-EP-241
  }}}%
\end{picture}%
How to extract the ``Abelian'' part of double-winding Wilson loop}

\ShortTitle{How to extract the ``Abelian'' part of double-winding Wilson loop}

\author{\speaker{Ryutaro Matsudo}\\
  Department of Physics, 
  Faculty of Science and Engineering, 
  Chiba University, Chiba 263-8522, Japan\\
    E-mail: \email{afca3071@chiba-u.jp}}

\abstract{It is known that the naive Abelian Wilson loop defined by the Abelian projection cannot reproduce the correct behavior of the double-winding Wilson loop. It is also known that the naive Abelian Wilson loop cannot reproduce the correct behavior of the Wilson loops in higher representations, but this problem was recently solved by using the redefined "Abelian" Wilson loop.
In this talk, we will give another reason why this redefined "Abelian" Wilson loop behaves correctly, and by following the same line of the argument, we will propose redefined "Abelian" double-winding Wilson loop which is considered to behave correctly.}

\FullConference{37th International Symposium on Lattice Field Theory - Lattice2019\\
		16-22 June 2019\\
		Wuhan, China}

\begin{document}

\section{Introduction}
The true mechanism of quark confinement is not known. One of the promising candidate is \textit{the dual superconductivity picture} \cite{dualsuper}.
In this picture, the QCD vacuum is considered as a dual superconductor, that is the electromagnetic dual of ausual superconductor.
Then the electric flux between a quark and an antiquark is squeezed into a tube due to the dual Meissner effect.
Thus the potential of the quark-antiquark pair linearly increases with the distance between the quark and the antiquark, which means quark confinement.
The ordinary superconductivity is the result of condensation of Cooper pairs and therefore we can suppose that condensation of magnetic monopoles have to occur for the QCD vacuum to be a dual superconductor.

The definition of magnetic monopoles in non-Abelian gauge theories is not so trivial.
The most famous way to define magnetic monopoles is \textit{the Abelian projection} \cite{Hooft81}.
In this method, we partially fix the gauge so that the symmetry corresponding to the diagonal transformations is left intact.
Then the diagonal component of the gauge field is identified with the Abelian gauge field and a magnetic monopoles is defined as a Dirac monopole.
It was numerically checked that magnetic monopoles defined by the Abelian projection dominantly contribute to the string tension between a quark and an antiquark as follows.
Firstly, it was investigated whether the diagonal component of the gauge field, that is used to define magnetic monopoles, reproduces the string tension between a quark and antiquark.
The contribution to the string tension from the diagonal component of the gauge field is defined by using Abelian Wilson loops, that are defined by replacing the gauge field with the diagonal component of the gauge field in the definition of Wilson loops. It was confirmed in the lattice simulations that the string tension extracted from Abelian Wilson loops reproduces the original string tension in the $SU(2)$ pure gauge theory \cite{SY90} and in the $SU(3)$ pure gauge theory \cite{STW02}.
This is called \textit{the Abelian dominance of the string tension}.
Next, it was investigated whether the monopole contribution reproduces the string tension.
The contribution to the string tension from magnetic monopoles is defined by applying the DeGrand-Toussaint procedure \cite{DT80} to Abelian Wilson loops.
It was checked in the lattice simulations that the monopole contribution defined in this way reproduces the string tension in the $SU(2)$ pure gauge theory \cite{SS94} and in the $SU(3)$ pure gauge theory \cite{STW02}.
This is called \textit{the monopole dominance of the string tension}.

In this way, in order to extract the monopole contribution by applying Abelian projection, the theory need to be described by using only the diagonal component of the gauge field in the low energy region. However, there are operators whose behaviors are not reproduced by the operators defined by just replacing the gauge field with the diagonal component of the gauge field.
One of such operators is a Wilson loop in a higher representation \cite{DFGO96}.
Naive Abelian Wilson loops in higher representations do not reproduce the behavior of the Wilson loops in higher representation.
We can avoid this problem if we find an ``Abelian'' operator that reproduces the behavior of the original operator in the low energy region.
Here ``Abelian'' means that the operator is defined by using only the diagonal component of the gauge field in the partially fixed gauge.
There is no reason why we consider that the Abelian operator must be defined by just replacing the gauge field with the diagonal component of the gauge field naively.
Such Abelian operators are found firstly for Wilson loops in the adjoint representation of $SU(2)$ \cite{Poulis96}, and then for Wilson loops in an arbitrary representation of an arbitrary gauge group \cite{MSKK19}.
The numerical simulations confirm this claim for Wilson loops in the adjoint representation of $SU(2)$ \cite{Poulis96,MSKK19} and in the adjoint representation and the sextet representation of $SU(3)$ \cite{MSKK19}.
Another operator whose behavior is not reproduced by the operator defined by just replacing the gauge field with the diagonal component of the gauge field is \textit{a double-winding Wilson loop} in the $SU(2)$ pure gauge theory \cite{GH15}.
A double-winding Wilson loop is the Wilson loop defined on the contour which consists of two coplanar loops, one of which lies entirely on the minimal area of the other loop. 
In this talk, we propose an Abelian operator that would reproduce the behavior of the double-winding Wilson loops.

\section{Abelian projection}
In this section, we give a brief review of the Abelian projection.
Frequently magnetic monopoles are defined by using the Abelian projection.
We consider $SU(N)$ gauge theories for simplicity.
In this method, link variables $U_l\in SU(N)$ are factorized into Abelian link variables $u_l\in U(1)^{N-1}$ and the off-diagonal parts $U^{\mathrm{off}}_l$ in a fixed gauge as
\begin{align}
  U_l = u_lU^{\mathrm{off}}_l.
\end{align}
An Abelian link variable $u_l$ is defined as the diagonal element of the gauge group that maximizes
\begin{align}
  \operatorname{Re}\tr (u_l U_l^\dag),
\end{align}
for each link.
This means that an Abelian link variable is the best diagonal approximation to a link variable.
This decomposition is performed in a fixed gauge.
The MA gauge where the functional
\begin{align}
  \sum_{l} \sum_{i=1}^{N-1} \tr(U_lH_iU_l^\dag H_i),
\end{align}
is maximized is frequently used.
Here $H_i$ is a Cartan generator.
The magnetic current is defined by applying the DeGrand-Touissaint procedure to this Abelian link variables.

\section{The Haar-measure-corrected Wilson loop}
In this section, we firstly reconsider single-winding Wilson loops in higher representations and then we extend the argument in order to consider double-winding Wilson loops.
It is known that the string tensions extracted from the Wilson loops in higher representations cannot be reproduced when the Abelian projection is applied naively.
To avoid this problem, in \cite{MSKK19}, ``the highest weight part of the Abelian Wilson loop'' is proposed and it is numerically checked that the string tensions of several higher representations are reproduced by using this operator.
In the following, we introduce another operator that reproduces the string tension in higher representations and give the reason why this operator reproduces the correct string tension \cite{MSKK18}.
Then we show this operator reproduces the correct string tension by assuming that the highest weight part of the Abelian Wilson loop reproduces the correct string tension.
Next, we modify the operator to reproduce the behavior of double-winding Wilson loops, whose contour consists of two coplanar loops, one of which lies entirely on the minimal area of the other loop.

Now we consider $SU(2)$ gauge theory for simplicity.
Let $W^{\mathrm{Abel}}(C) = \operatorname{diag}(e^{i\theta_C},e^{-i\theta_C})$ be the untraced Abelian Wilson loop defined on a loop C.
Then it is proposed in \cite{MSKK18} that the following operator reproduces the correct string tension in the spin-$J$ representation:
\begin{align}
 2\sin^2 \theta_C \tr_JW^{\mathrm{Abel}}(C), \label{haar_op}
\end{align}
where $\tr_J$ denotes the trace in the spin-$J$ representation, i.e.,
\begin{align}
 \tr_JW^{\mathrm{Abel}}(C) := e^{-2iJ\theta_C} + e^{-2i(J-1)\theta_C} + \cdots + e^{2iJ\theta_C}.
\end{align}
We call the operator (\ref{haar_op}) as \textit{the Haar-measure-corrected Abelian Wilson loop} because the factor $2\sin^2\theta_C$ corresponds to the difference of the Haar measure between $SU(2)$ and $U(1)$.

The reason why we consider this operator behaves correctly is as follows. Firstly we write the expectation value of a Wilson loop and an Abelian Wilson loop by using distribution function of the untraced Wilson loop and that of the Abelian Wilson loop as
\begin{align}
  \braket{\tr_JW(C)} &= \int_{SU(2)} dW\, P(W;C)\tr_J W \label{dist1}\\
  \braket{\tr_JW^{\mathrm{Abel}}(C)}&=\int_{U(1)}dw\, P^{\mathrm{Abel}}(w;C) \tr_J w,\label{dist2}
\end{align}
where $dW$ is the Haar measure on $SU(2)$, $dw$ is the Haar measure on $U(1)$, and the distribution functions are defined as
\begin{align}
  P(W;C) &:= \int DU\, \delta_{SU(2)}(W-\prod_{l\in C}U_l)e^{-S[U]}, \\
  P^{\mathrm{Abel}}(w:C) &:= \int DU \,\delta_{U(1)}(w-\prod_l u_l) e^{-S[U]}, 
\end{align}
where $\delta_{SU(2)}$ and $\delta_{U(1)}$ are the delta functions on $SU(2)$ and $U(1)$ respectively and $S[U]$ is the action.
Because $P(W;C)$ is a class function, i.e., $P(W) = P(gWg^{-1}), \forall g\in SU(2)$, it can be written as a function of the eigenvalues $e^{\pm i\theta}$ of $W$ and then
\begin{align}
  P(W;C) = P(\theta;C).
\end{align}
Then Eqs.\ (\ref{dist1}) and (\ref{dist2}) reduce to
\begin{align}
  \braket{\tr_JW(C)} &= \frac1\pi\int_0^\pi d\theta \,2\sin^2\theta P(\theta;C) \sum_{k=0}^J e^{2i(k-J)\theta}, \label{dist1'}\\
  \braket{\tr_JW^{\mathrm{Abel}}(C)}&=\frac1{2\pi}\int_0^{2\pi}d\theta\, P^{\mathrm{Abel}}(\theta;C) \sum_{k=0}^J e^{2i(k-J)\theta} \label{dist2'}.
\end{align}
If $P(\theta;C)$ and $P^{\mathrm{Abel}}(\theta:C)$ behave similarly, the difference between Eq.\ (\ref{dist1'}) and Eq.\ (\ref{dist2'}) is only the factor $2\sin^2\theta$.
  Therefore, we guess from this that if the operator including the difference of the measure is modified as
  \begin{align}
  \braket{2\sin^2\theta_C\tr_JW^{\mathrm{Abel}}(C)} = \frac1\pi\int_0^\pi d\theta \,2\sin^2\theta P^{\mathrm{Abel}}(\theta;C) \sum_{k=0}^J e^{2i(k-J)\theta},
  \end{align}
the correct behavior is obtained.

%\begin{figure}[t]
%\centering
%\includegraphics[width=0.2\hsize]{contour_double.pdf}
%\caption{The contour of a double-winding Wilson loop. $C_1$ and $C_2$ are coplanar and $C_1$ lies entirely on the minimal area of $C_2$}
%\label{fig1}
%\end{figure}

It has been already checked to some extent that the Haar-measure-corrected Abelian Wilson loop behaves correctly.
This is because, by using an operator relation, we can relate the Haar-measure-corrected Abelian Wilson loop and the highest-weight part $\widetilde W_J^{\mathrm{Abel}}(C)$ of the Abelian Wilson loop, that is defined as
\begin{align}
  \widetilde W_J^{\mathrm{Abel}}(C) := \frac12(e^{2iJ\theta_C} + e^{-2iJ\theta_C}),
\end{align}
and it has been numerically shown that this behaves similarly to the original Wilson loop in the spin-$J$ representation
\begin{align}
  \braket{\widetilde W_J^{\mathrm{Abel}}(C)} \sim \braket{\tr_JW(C)}
\end{align}
when $J=1$ in the previous studies \cite{Poulis96,MSKK19}.
Here the symbol ``$\sim$'' means that the both sides decrease exponentially with the area surrounded by $C$ at approximately the same rate.
The operator relation, which relates the highest-weight part of the Abelian Wilson loop and the Haar-measure-corrected Abelian Wilson loop, is
\begin{align}
 2\sin^2\theta \tr_JW^{\mathrm{Abel}} = \widetilde W_J^{\mathrm{Abel}} - \widetilde W_{J+1}^{\mathrm{Abel}}
\end{align}
By taking the average we obtain 
\begin{align}
 \braket{2\sin^2\theta \tr_JW_J^{\mathrm{Abel}}} &= \braket{\widetilde W_J^{\mathrm{Abel}}} - \braket{\widetilde W_{J+1}^{\mathrm{Abel}}}\notag\\
 &\sim c_1e^{-\sigma_J A} + c_2 e^{-\sigma_{J+1} A} \notag\\
 &\sim e^{-\sigma_J A}. 
\end{align}
where $c_1$ and $c_2$ are constants, $A$ is the minimal area surrounded by $C$, $\sigma_J$ is the string tension in the spin-$J$ representation and the second similarity is because $\sigma_{J+1}>\sigma_J$.

Next we consider double-winding Wilson loops.
In the $SU(2)$ pure gauge theory, it was numerically checked that double-winding Wilson loops obey the difference-of-areas law, i.e.,
\begin{align}
  \braket{\tr (W(C_1)W(C_2))} \sim e^{-\sigma(A_2-A_1)},
\end{align}
where $W(C_i)$ is the untraced Wilson loop whose contour is $C_i$ \cite{GH15}.
Here $C_1$ and $C_2$ are coplanar loops with the minimal areas $A_1$ and $A_2$ respectively and $C_1$ lies entirely on the minimal area of $C_2$.
The symbol ``$\sim$'' means that the both sides decrease exponentially with the areas $A_1$ and $A_2$ at approximately the same rate.
Recently, it has been discussed that naive Abelian Wilson loops cannot reproduce this behavior of double-winding Wilson loops \cite{GH15}.
For double-winding contours, the straightforward modification of the operator is
\begin{align}
 2\sin^2\theta_12\sin^2\theta_2 W^{\mathrm{Abel}}(C_1)W^{\mathrm{Abel}}(C_2), \label{haar_c_w}
%\sim \braket{[W(C_1)]^a{}_b[W(C_2)]^c{}_d}, \label{haar_c_w}
\end{align}
where the untraced Abelian Wilson loop on a contour $C_i$ is parametrized as
\begin{align}
  W^{\mathrm{Abel}}(C_i) = \operatorname{diag}(e^{i\theta_i},e^{-i\theta_i}).
\end{align}
We call the operator (\ref{haar_c_w}) as \textit{the Haar-measure-corrected Abelian double-winding Wilson loop}.

\section{Behavior of the proposed operators on double-winding contour}

Let us estimate the average of the Haar-measure-corrected Abelian double-winding Wilson loops.
Let $C_1$ and $C_2$ be two coplanar loops, where $C_1$ lies entirely on the minimal area of $C_2$,
and $\operatorname{diag}(e^{i\theta_i},e^{-i\theta_i})$ be the untraced Abelian Wilson loop on a contour $C_i$.
Then the operator we consider is
\begin{align}
  2\sin^2\theta_12\sin^2\theta_2\cos(\theta_1+\theta_2).
\end{align}

This operator is decomposed as
\begin{align}
  &2\sin^2\theta_12\sin^2\theta_2\cos(\theta_1+\theta_2)\notag\\
  &= -\cos(\theta_2-\theta_1) + \frac54\cos(\theta_1+\theta_2)
  +\frac14\cos(3(\theta_1+\theta_2)) \notag\\& + \frac14\cos(3\theta_2-\theta_1)
  +\frac14\cos(\theta_2-3\theta_1) -\frac12\cos(3\theta_1+\theta_2)
  -\frac12\cos(\theta_1+3\theta_2). \label{decom}
\end{align}
In the following we estimate each terms.

The first term gives
\begin{align}
  \braket{\cos(\theta_2-\theta_1)} \sim e^{-\sigma_{\mathrm{fund}}(A_2-A_1)}.
\end{align}
This is because the first terms is the Wilson loop with the contour that winds once around $C_2$ and once around $C_1$ in the inverse direction.
This is usual single-winding Wilson loop with the minimal area $A_2-A_1$.
This gives the difference-of-areas law.
The other terms decrease faster as shown below.

%\begin{figure}[t]
%\centering
%\includegraphics[width=0.6\hsize]{./contour_double_identical+difference.pdf}
%\caption{The second term in Eq.\ (\ref{decom}) factorizes in to two parts according to the assumption (\ref{assum}). $C_1^2$ and $C_2C_1^{-1}$ are non-intersecting coplanar loops and therefore we can use the assumption.}
%\label{fig2}
%\end{figure}
In order to estimate the average of other terms, we assume factorization for non-intersecting coplanar Abelian Wilson loops as
\begin{align}
  \braket{[W^{\mathrm{Abel}}(\widetilde C_1)]^a{}_b[W^{\mathrm{Abel}}(\widetilde C_2)]^c{}_d} \sim \braket{[W^{\mathrm{Abel}}(\widetilde C_1)]^a{}_b}\braket{[W^{\mathrm{Abel}}(\widetilde C_2)]^c{}_d}, \label{assum}
\end{align}
where $\widetilde C_1$ and $\widetilde C_2$ are non-intersecting coplanar loops. %and the symbol ``$\sim$'' means that the both sides decrease exponentially with the areas $A_1$ and $A_2$ at approximately the same rate.
This assumption is not so strange because this is consistent with the usual area law and the two-dimensional gauge theory.
By using this assumption we can factorize the terms in Eq.\ (\ref{decom}). For example, the second term is estimated as
\begin{align}
  &\braket{\cos(\theta_1+\theta_2)} = \braket{e^{i\theta_1}e^{i\theta_2}} = \braket{e^{2i\theta_1}e^{i(\theta_2-\theta_1)}} \sim \braket{e^{i2\theta_1}}\braket{e^{i(\theta_2-\theta_1)}} \notag\\
  &\sim e^{- \sigma_{\mathrm{adj}} A_1 -\sigma_\mathrm{fund} (A_2-A_1)},
\end{align}
where the first equality is because of the symmetry corresponding to the Weyl reflection, the first similarity is because of the assumption (\ref{assum}) and the last similarity is because the operator $e^{i2\theta_1}$ is actually the highest weight part in the adjoint representation. 
%See Fig.\ \ref{fig2}.
As we explained before, the numerical simulations confirms that the highest weight part in the adjoint representation of $SU(2)$ gives approximately the same value of the correct string tension for the adjoint sources.
Similarly we estimate the other terms as
\begin{align}
  &\braket{\cos(3(\theta_1+\theta_2))} \sim e^{- \sigma_{[J=3]}A_1 - \sigma_{[J=3/2]}(A_2-A_1)},\qquad
  &&\braket{\cos((3\theta_2-\theta_1))} \sim e^{- \sigma_{\mathrm{adj}}A_1 - \sigma_{[J=3/2]}(A_2-A_1)},\notag\\
  &\braket{\cos((\theta_2-3\theta_1))} \sim e^{- \sigma_{\mathrm{adj}}A_1 - \sigma_{\mathrm{fund}}(A_2-A_1)},\qquad
  &&\braket{\cos(3\theta_1+\theta_2)} \sim e^{- \sigma_{[J=2]}A_1 - \sigma_{\mathrm{fund}}(A_2-A_1)},\notag\\
  &\braket{\cos(\theta_1+3\theta_2)} \sim e^{- \sigma_{[J=2]}A_1 - \sigma_{[J=3/2]}(A_2-A_1)}.
\end{align}
Here we assume additionally that the highest weight part gives approximately the same value of the correct string tension also in higher representations other than the adjoint one.

Thus the terms other than $-\braket{\cos(\theta_1-\theta_2)}$ decrease exponentially with $A_1$ when $A_2-A_1$ is held constant, and therefore for sufficiently large $A_1$ we obtain 
\begin{align}
  &\braket{2\sin^2\theta_12\sin^2\theta_2\cos(\theta_1+\theta_2)}\notag\\
%  &= -\braket{\cos(\theta_2-\theta_1)} + \frac54\braket{\cos(\theta_1+\theta_2)}
%  +\frac14\braket{\cos(3(\theta_1+\theta_2))} \notag\\& + \frac14\braket{\cos(3\theta_2-\theta_1)}
%  +\frac14\braket{\cos(\theta_2-3\theta_1)} -\frac12\braket{\cos(3\theta_1+\theta_2)}
%  -\frac12\braket{\cos(\theta_1+3\theta_2)} \\
  &\sim e^{\sigma_{\mathrm{fund}}(A_2-A_1)}.
\end{align}
This is nothing but the difference-of-areas law.

\section{Conclusion}
We propose a way to extract ``Abelian'' part which could apply to Wilson loops in higher representations and double-winding Wilson loops.
Because of the operator relation between the proposed operators and the highest-weight part of Abelian Wilson loops, the proposed operators reproduce the correct area-law behavior of Wilson loops in higher representation.
Under an assumption, we estimate the average of the proposed operators defined on double winding contours by using Abelian link variables, which give the difference-of-areas law.

\section*{Acknowledgement}
This work was supported by Grant-in-Aid for JSPS Research Fellow Grant Number 17J04780.

\end{document}